\documentclass[aps,reprint,amsmath,amssymb,superscriptaddress,prd]{revtex4-2}
\usepackage{graphicx}
\usepackage{subfigure}
\usepackage{bm}
\usepackage{mathrsfs,ae,dsfont}
\usepackage{color}
\usepackage{hyperref}
\hypersetup{hypertex=true,
colorlinks=true,
linkcolor=blue,
anchorcolor=blue,
citecolor=blue,
filecolor=blue,
urlcolor=blue,
citecolor=blue}

\begin{document}
\title{On the nonlinearity of Four-Dimensional Conformal Transformations in spinor representation}

\author{Zhi-Peng Wang}\email{wangzp@nenu.edu.cn}
\author{X. X. Yi}
\affiliation{Center for Quantum Sciences and School of Physics, Northeast Normal University, Changchun 130024, China}

\author{Hai-Jun Wang}\email{hjwang@jlu.edu.cn}
\affiliation{Center for Theoretical Physics and College of Physics, Jilin University, Changchun 130012, China}

\date{\today}

\begin{abstract}
    The nonlinearity of the conformal group is an essential factor that ruins the global conformal invariance for interacting material fields. In this paper we attempt to track such nonlinearity from spacetime transformations to spinor representations. To this end we rederive the spinor representation by generalizing the linear fractional transformation from two dimensions to four dimensions via replacing complex numbers with biquaternions. To check the effect of the nonlinearity we apply the translations and special conformal transformations (SCTs) to Dirac spinors in certain interactions. These two transformations do not lead to nonlinear terms in Yukawa term, but do in vector-spinor interaction. And the nonlinear terms would definitely cause $CP$ violation. 
\end{abstract}
    \maketitle

\section{introduction}

The study of conformal group in physics has been updating ever since Bateman and
Cunningham discovered \cite{01,02} that Maxwell's equations are invariant
under conformal transformations in 1910. The conformal group is composed of
conformal transformations with Poincar\'e group as its subgroup. Additional
to translations and rotations, the conformal group includes dilations and
special conformal transformations (SCTs). The dilation, also known as
scaling transformation, is widely used in different renormalization schemes 
\cite{03}. The inversion is also a conformal transformation, but it is
somehow equivalent to the SCT which is constructed through the combination
of a inversion, a translation, and a second inversion. The properties of the
SCT are particularly distinctive, as they can be locally viewed as the
combination of a rotation and dilation with coordinate-dependent
coefficients, which illustrates the nonlinear nature of the SCT. So far, it remains elusive for us that
how the nonlinearity manifests itself in different representations of
conformal group, for instance in spinor representations.

It is well known that few actions and equations maintain conformal 
invariance due to the nonlinear nature of the SCT, for instance, the
acceleration equation in special relativity \cite{04,05} and the action of
the photon field \cite{06}. The nonlinear properties of the SCT have been
studied in previous literatures \cite{07,08}, where it seems transforming
the trajectory of a particle moving in a straight line into two particles
moving along a hyperbola. Another work concerns conformal transformations in
the four-dimensional electrodynamics \cite{09}, where the two particles
moving along the hyperbola are interpreted by the Born solutions when the
selected electric field is Coulomb field. Recently, in the conformal
bootstrap method \cite{10,11}, the nonlinearity of the SCT has been
manifested by the finite transformations of ``primary operator'', leading to
the determination of the critical exponent of the three-dimensional Ising
model \cite{12}. Even more, the conformal group has been schematically applied
to conformal quantum mechanics, chaos and cosmology \cite{13,14,15,16}, where the
conformal group is reduced to the SL(2,R) group.

Up to date, the conformal transformations for scalar and vector have been
extensively studied. Whereas the study for spinor \cite{17,18,19} remain not
adequate, for instance how to apply them to Dirac spinors. Consequently, 
two issues arise while regarding the four-dimensional SCT. The first 
issue concerns the nonlinear effects of the SCT in different representations 
which have not been explicitly compared. The second issue relates to 
the derivation of spinor representations of the conformal group \cite{17}, 
i.e. how to connect the spinor representation with four-dimensional spacetime 
transformations Eqs. \eqref{Def_CT}. This connection should not be confused with
progress in deriving the four-dimensional conformal group via six-dimensional 
``isotropic-vector" \cite{20,21,22} (the vector components meet $x_1^2+x_2^2+x_3^2+x_4^2+x_5^2+x_6^2=0$). In that case, spinor 
representations of the four-dimensional conformal group can also be derived. 
However, the coordinate transformations in six dimensions are not identical 
to those Eqs. \eqref{Def_CT} in four dimensions.

In two dimensions, conformal transformations can be associated to linear 
fractional transformations (projective transformations, Mobius transformations) \cite{23}. Therefore to
address the two issues mentioned above, we extend linear fractional
transformations from two dimensions to four dimensions, using biquaternions  \cite{24}
to replace complex numbers. In a recent work \cite{25},
hyperquaternions were employed to derive the six-dimensional coordinate
transformations and four-dimensional generators of the conformal group.
Whereas to explicitly manifest the connection between four-dimensional
spacetime transformations and the spinor representations, biquaternions prove to be convenient \cite{26}. Our results show
that both translations and SCTs have nonlinear effects while acting on the 
vector-spinor interaction, however while they acting on the Yukawa term, the nonlinearity vanishes. At the early moments of universe, the acceleration should be expressed in a transformation manner merely by translations and SCTs, which we define
in this paper as the physical translations.
Such physical translations would cause somehow the violation of charge conjugation parity ($CP$).

The remainder of the paper is structured as follows. In Sec. \ref{Sec_2}, we
present the basic definition of the conformal transformation. In Sec. \ref{Sec_3}, we
associate the coordinate transformation with the spinor representation of
the conformal group via generalized linear fractional transformations. In
Sec. \ref{Sec_4}, we apply conformal transformations to Dirac spinors. In
Sec. \ref{Sec_5}, we discuss our results.
\section{The Basis of Conformal Transformations}\label{Sec_2}
In this section, we introduce conformal transformations, with a particular focus on the properties of SCTs.  In Minkowski space with the metric $g^{\mu\nu}=\text{diag}(1,-1,-1,-1)$, finite conformal transformations are defined to be 
\begin{subequations}\label{Def_CT}
    \begin{align}
        \textrm{(translation)} \enspace &x^{\mu\prime}=  x^{\mu}+a^{\mu},\\
       \label{Def_R} \textrm{(rotation)} \enspace &x^{\mu\prime}=  \Lambda_{\enspace \nu}^{\mu}x^{\nu},\\
        \label{Def_D} \textrm{(dilation)} \enspace &x^{\mu\prime}=  e^\alpha x^{\nu},\\
         \label{Def_SCT} \textrm{(SCT)} \enspace &x^{\mu\prime}=  \frac{x^{\mu}-b^{\mu}x^{2}}{(1-2b\cdot x+b^{2}x^{2})},
    \end{align}
\end{subequations}
among which the SCT consists of a series of transformations,
\begin{subequations}\label{DEF_I_T_I}
    \begin{align}
       \label{Def_I} x^{\mu\prime}=&\frac{x^{\mu}}{x^{2}},\\
       \label{I_T} x^{\mu\prime\prime}=&\frac{x^{\mu}}{x^{2}}-b^{\mu},\\
       \label{I_T_I}  x^{\mu\prime\prime\prime}=&\frac{\frac{x^{\mu}}{x^{2}}-b^{\mu}}{(\frac{x^{\nu}}{x^{2}}-b^{\nu})(\frac{x_{\nu}}{x^{2}}-b_{\nu})},
    \end{align}
\end{subequations}
where Eq. \eqref{Def_I} is defined as inversion, Eqs. \eqref{I_T} and \eqref{I_T_I} are known as translation and second inversion. Obviously, inversions are the most important operations for finite conformal transformations. However while constructing infinitesimal conformal transformations, inversions are replaced by SCTs. Unlike translation, rotation, and dilation, SCTs are nonlinear transformations. The generators of the conformal group can be derived through infinitesimal transformations. The explicit forms of these generators are
\begin{subequations}
    \begin{align}
        P_{\mu}= & -i\partial_{\mu},\\
        M_{\mu\nu}= & -i(x_{\nu}\partial_{\mu}-x_{\mu}\partial_{\nu}),\\
        D= & -ix^{\mu}\partial_{\mu},\\
        K_{\mu}= & -i(2x_{\mu}x^{\nu}\partial_{\nu}-x^{2}\partial_{\mu}),
    \end{align}
\end{subequations}
where, $P_\mu$, $M_{\mu\nu}$, $D$, and $K_\mu$ are respectively generators of translations, rotations , dilations, and SCTs. A SCT locally looks like the combination of a dilation $\Omega(x)$ and a rotation $R_{\enspace\nu}^{\mu}(x)$, which is shown by the Jacobian of the SCT \cite{27} 
\begin{equation}\label{Jac}
    \frac{\partial x^{\mu\prime}}{\partial x^{\nu}}=\Omega(x)R_{\enspace\nu}^{\mu}(x),
\end{equation}
where  $\Omega(x)$ and $R_{\enspace\nu}^{\mu}(x)$ are  
\begin{equation}\label{Jac_D_M}
    \begin{aligned}
        \Omega(x) =& (1-2b\cdot x+b^{2}x^{2})^{-1},\\
        R_{\enspace\nu}^{\mu}(x)=&\delta_{\nu}^{\mu}+\Omega(x)[2(b_{\nu}x^{\mu}-b^{\mu}x_{\nu})+4b\cdot xb^{\mu}x_{\nu}\\&-2(b^{\mu}b_{\nu}x^{2}+b^{2}x^{\mu}x_{\nu})].
    \end{aligned}
\end{equation}
According to Eq. \eqref{Jac_D_M}, we observe that the powers of translation parameter (coefficient) $b^\mu$ and coordinate $x^\mu$ are the same. Thus, in the spinor representation, we shall regard the power of $b^\mu$ as the degree of nonlinearity of the conformal transformations, more details as following Eq. \eqref{Int_SCT} .

\section{Rederive the Spinor Representation of Conformal Group}\label{Sec_3}

In this section, we will generalize the linear fractional transformation $\Lambda=\left(\begin{array}{cc}
a & b\\
c & d
\end{array}\right)$, by using biquaternions in place of complex numbers (Here the letters a, b, c, d have nothing to do with those in last section, we don't stress that henceforth unless confusion appears.). By such
generalization we extend the conformal transformations from two-dimensions to
four-dimensions, meanwhile we obtain the mapping between spacetime transformations and
spinor representations of the conformal group.

\subsection{Spinor Representation in two dimension}

Now we introduce the linear fractional transformation by which conformal transformations can be associated with matrices. And using these
matrices we define a representations of conformal group. It is well known
that a the linear fractional transformation can be associated to the matrix
$\Lambda=\left(\begin{array}{cc}
a & b\\
c & d
\end{array}\right)$ \cite{23}, concretely as 
\begin{equation}\label{Def_LFT}
    z^{\prime}=\frac{az+b}{cz+d},
\end{equation}
where $a$, $b$, $c$, $d$ and $z=x+iy$ are complex numbers. The mappings \eqref{Def_LFT} correspond to conformal transformations \cite{28}, among which the translation in the $x$ direction can be associated with $\Lambda_{P_x}=\left(\begin{array}{cc} 
    1 & a\\
    0 & 1
\end{array}\right)$. Substituting this matrix into Eq. \eqref{Def_LFT}, we can get 
\begin{equation}
    z^{\prime}=z+a=x+a+iy.
\end{equation}
The corresponding coordinate transformation then is 
\begin{equation}
    \begin{aligned}
        x^{\prime}=&x+a,\\
        y^{\prime}=&y.
    \end{aligned}
\end{equation} 
Similarly, translation in the $y$ direction can be associated with $\Lambda_{P_y}=\left(\begin{array}{cc}
        1 & ai\\
        0 & 1
        \end{array}\right)$.
The corresponding coordinate transformation is $x^{\prime}=x$, $ y^{\prime}=y+a$.
The rotation can be associated with
$\Lambda_{M}=\left(\begin{array}{cc}
        e^{\frac{i\theta}{2}} & 0\\
        0 & e^{-\frac{i\theta}{2}}
        \end{array}\right)$. 
The corresponding coordinate transformation is
\begin{equation}
    \begin{aligned}
        x^{\prime}= &(\cos\theta)x+(\sin\theta)y,\\
        y^{\prime}=& -(\sin\theta)x+(\cos\theta)y.
    \end{aligned}
\end{equation}
The dilation can be associated with
$\Lambda_{D}=\left(\begin{array}{cc}
        e^{\frac{\alpha}{2}} & 0\\
        0 & e^{-\frac{\alpha}{2}}
        \end{array}\right).$  
The corresponding coordinate transformation is 
\begin{equation}
    x^{\prime}=  e^{\alpha}x,\enspace y^{\prime}=  e^{\alpha}y.
\end{equation}
Similar to Eq. \eqref{DEF_I_T_I}, SCTs are derived 
\begin{align}
    \label{K_x} \Lambda_{K_{x}}= &\Lambda_{I_{x}}\Lambda_{P_{x}}\Lambda_{I_{x}}
        =\left(\begin{array}{cc}
        1 & 0\\
        -b & 1
        \end{array}\right),\\
    \label{K_y} \Lambda_{K_{y}}= & \Lambda_{I_{y}}\Lambda_{P_{y}}\Lambda_{I_{y}}
    =\left(\begin{array}{cc}
    1 & 0\\
    bi & 1
    \end{array}\right),
\end{align}
where $\Lambda_{I_{x}}$ is the $x$-directional inversion,  and $\Lambda_{I_{y}}$ $y$-directional, which can be expressed as
\begin{equation}\label{2dI}
   \Lambda_{I_{x}} =\left(\begin{array}{cc}
0 & 1\\
1 & 0
\end{array}\right),\enspace 
\Lambda_{I_{y}}=\left(\begin{array}{cc}
0 & i\\
-i & 0
\end{array}\right),
\end{equation}
in which $\Lambda_{I_{x}}$ and $\Lambda_{I_{y}}$ corresponding coordinate transformation are 
\begin{equation}\label{I}
    \begin{aligned}
        x_{I_{x}}^{\prime}=\frac{x}{x^{2}+y^{2}},\enspace & y_{I_{x}}^{\prime}=\frac{-y}{x^{2}+y^{2}},\\
x_{I_{y}}^{\prime}=\frac{-x}{x^{2}+y^{2}},\enspace & y_{I_{y}}^{\prime}=\frac{y}{x^{2}+y^{2}}.
    \end{aligned}
\end{equation}
The SCT in the $x$ direction \eqref{K_x}  corresponding coordinate transformation is
\begin{equation}\label{SCT_x}
    \begin{aligned}
    x^{\prime}=&  \frac{x-b(x^{2}+y^{2})}{1-2bx+b^{2}(x^{2}+y^{2})},  \\
   y^{\prime}= &\frac{y}{1-2bx+b^{2}(x^{2}+y^{2})}.
    \end{aligned}
\end{equation}
The SCT in the $y$ direction  \eqref{K_y} corresponding coordinate transformation is
\begin{equation}\label{SCT_y}
    \begin{aligned}
        x^{\prime}= & \frac{x}{1-2by+b^{2}(x^{2}+y^{2})},\\
y^{\prime}= & \frac{y-b(x^{2}+y^{2})}{1-2by+b^{2}(x^{2}+y^{2})}.
    \end{aligned}
\end{equation}
By the above Eqs. \eqref{SCT_x} and \eqref{SCT_y}, we restore the coordinate transformations of SCTs in Eq. \eqref{Def_CT}. In such representation, the corresponding generators of conformal transformations in two dimensions yield
\begin{equation}
    \begin{aligned}
        P_{x}=\left(\begin{array}{cc}
        0 & i\\
        0 & 0
        \end{array}\right),\enspace&
        P_{y}=\left(\begin{array}{cc}
        0 & -1\\
        0 & 0
        \end{array}\right),\\
        M=\left(\begin{array}{cc}
        -\frac{1}{2} & 0\\
        0 & \frac{1}{2}
        \end{array}\right),\enspace&
        D=\left(\begin{array}{cc}
        \frac{i}{2} & 0\\
        0 & -\frac{i}{2}
        \end{array}\right),\\
        K_{x}=\left(\begin{array}{cc}
        0 & 0\\
        -i & 0
        \end{array}\right),\enspace&
        K_{y}=\left(\begin{array}{cc}
        0 & 0\\
        -1 & 0
        \end{array}\right).
    \end{aligned}
\end{equation}
And their commutation relations can be easily examined, 
\begin{equation}
    \begin{aligned}
    [D,P_{x}]=iP_{x},\enspace & [D,P_{y}]=iP_{y},\\
[D,K_{x}]=-iK_{x},\enspace & [D,K_{y}]=-iK_{y},\\
[K_{x},P_{x}]=2iD,\enspace & [K_{x},P_{y}]=-2iM,\\
[K_{y},P_{x}]=2iM,\enspace & [K_{y},P_{y}]=2iD,\\
[P_{x},M]=iP_{y},\enspace & [P_{y},M]=-iP_{x},\\
[K_{x},M]=iK_{y},\enspace & [K_{y},M]=-iK_{x}.
    \end{aligned}
\end{equation} 

\subsection{Spinor Representation in four dimension}

Now let's turn to the four-dimensional linear fractional transformation. We would
replace the complex number with the biquaternion \cite{24}, which is defined to be  
\begin{equation}
        q=t+x(h\boldsymbol{i})+y(h\boldsymbol{j})+z(h\boldsymbol{k}),
\end{equation} 
where $h$ is the imaginary unit, and $h$, $\boldsymbol{i}$, $\boldsymbol{j}$,  $\boldsymbol{k}$ satisfy the multiplication law \cite{29}  
\begin{align*}
    & \boldsymbol{i}^{2}=\boldsymbol{j}^{2}=\boldsymbol{k}^{2}=-1,\\
    & \boldsymbol{i}\boldsymbol{j}=\boldsymbol{k},\enspace\boldsymbol{j}\boldsymbol{k}=\boldsymbol{i},\enspace\boldsymbol{k}\boldsymbol{i}=\boldsymbol{j},\\
    & h\boldsymbol{i}=\boldsymbol{i}h,\enspace h\boldsymbol{j}=\boldsymbol{j}h,\enspace h\boldsymbol{k}=\boldsymbol{k}h,
\end{align*}
the number $1$ playing the role of unity element. Then the denominator in Eq. \eqref{Def_LFT} can be expressed by the inverse of biquaternions, which means the generalized linear fractional transformation can be written as 
\begin{equation}\label{Def_LFT_Q}
    q^{\prime}=(aq+b)(cq+d)^{-1},
\end{equation}
where $a$, $b$, $c$, and $d$ are biquaternions. In some literatures, this is also regarded as the definition of the linear fractional transformation \cite{19,30}. And the inverse of biquaternions \cite{24,31} is defined to be
\begin{equation}
    \begin{aligned}
        q^{-1}= (\bar{q}q)^{-1}\bar{q},
    \end{aligned}
\end{equation}
where quaternion conjugate $\bar{q}$ ($\boldsymbol{i},\boldsymbol{j},\boldsymbol{k}\rightarrow -\boldsymbol{i},-\boldsymbol{j},-\boldsymbol{k}$) and complex conjugate $q^\ast$ ($h\rightarrow -h$) are defined to be
\begin{equation}
    \bar{q}=q^\ast=t-x (h\boldsymbol{i})-y (h\boldsymbol{j})-z (h\boldsymbol{k}).
\end{equation}
Similarly, the generalized linear fractional transformation can be associated to the matrix 
$\Lambda=\left(\begin{array}{cc}
a & b\\
c & d
\end{array}\right)$. By using such replacement, translations turn out to be
\begin{equation}\label{Tral}
    \Lambda_{P}=\left(\begin{array}{cc}
        1 & a^{\mu}\sigma_{\mu}\\
        0 & 1
        \end{array}\right),
\end{equation} 
where $\sigma^{\mu}=(1,\sigma_i)$, $\bar{\sigma}^{\mu}=(1,-\sigma_i)$, $\sigma_i=-h\boldsymbol{i}, -h\boldsymbol{j}, -h\boldsymbol{k}$ and $1$ is the unit matrix. Substituting this matrix \eqref{Tral} into Eq. \eqref{Def_LFT_Q}, we can get 
\begin{equation}
    \begin{aligned}
        q^{\prime}= & (aq+b)(cq+d)^{-1}\\
        = & t+x(h\boldsymbol{i})+y(h\boldsymbol{j})+z(h\boldsymbol{k})+a^{\mu} {\sigma}_{\mu}\\
        = & t+x(h\boldsymbol{i})+y(h\boldsymbol{j})+z(h\boldsymbol{k})\\
         & +a^{0}+a^{1}{\sigma}_{1}+a^{2}{\sigma}_{2}+a^{3}{\sigma}_{3}\\
        = & t+x(h\boldsymbol{i})+y(h\boldsymbol{j})+z(h\boldsymbol{k})\\
         & +a_{t}+a_{x}(h\boldsymbol{i})+a_{y}(h\boldsymbol{j})+a_{z}(h\boldsymbol{k})\\
        = & t+a_{t}+(x+a_{x})(h\boldsymbol{i})\\
        &+(y+a_{y})(h\boldsymbol{j})+(z+a_{z})(h\boldsymbol{k}).    
    \end{aligned}
\end{equation}
One finds that the corresponding translations is
\begin{equation}
    \begin{aligned}
        t^{\prime}= & t+a_{t},\\
        x^{\prime}= & x+a_{x},\\
        y^{\prime}= & y+a_{y},\\
        z^{\prime}= & z+a_{z}.\\
    \end{aligned}
\end{equation} 
More generally, this can be written as
\begin{equation}
    x^{\mu\prime}= x^{\mu}+a^{\mu}.
\end{equation}
Similarly, the SCT \eqref{Def_SCT} can be derived analogous to Eqs. (\ref{K_x}, \ref{K_y}, \ref{SCT_x}, \ref{SCT_y}) 
\begin{equation}\label{SCT}
    \begin{aligned}
        \Lambda_{K}= &  \prod_{\alpha=0}^{3}\Lambda_{I_\alpha}\Lambda_{P_\alpha}\Lambda_{I_\alpha}\\
        = &\left(\begin{array}{cc}
            1 & 0\\
            -b^{\mu} \bar{\sigma}_{\mu} & 1
            \end{array}\right),
    \end{aligned}
\end{equation}
where $\Lambda_{I_{0}}$ is the temporal inversion, and $\Lambda_{I_{i}}$ is spatial inversions, which can be expressed as analogous to Eq. \eqref{2dI}
\begin{align}
    \label{I_0} \Lambda_{I_{0}}=&\left(\begin{array}{cc}
        0 & 1\\
        1 & 0
        \end{array}\right),\\
    \label{I_i} \Lambda_{I_{i}}=&\left(\begin{array}{cc}
        0 & h\sigma_{i}\\
        -h\sigma_{i} & 0
        \end{array}\right).
\end{align}
Correspondingly the rotation \eqref{Def_R} and dilation \eqref{Def_D} are respectively associated with 
\begin{equation}\label{R_D}
    \begin{aligned}
        \Lambda_{M}= &\left(\begin{array}{cc}
        e^{\frac{1}{2}\omega^{\mu\nu}\sigma_{\mu\nu}} & 0\\
        0 & e^{\frac{1}{2}\omega^{\mu\nu}\bar{\sigma}_{\mu\nu}}
        \end{array}\right),\\ 
        \Lambda_{D}= &\left(\begin{array}{cc}
        e^{\frac{\alpha}{2}} & 0\\
        0 & e^{-\frac{\alpha}{2}}
        \end{array}\right),
    \end{aligned}
\end{equation}
where $\sigma_{\mu\nu}$ and $\bar{\sigma}_{\mu\nu}$  is an antisymmetric tensor, which can analogously be expressed as
\begin{equation}
    \sigma^{\mu\nu}=\frac{1}{4}(\sigma^{\mu}\bar{\sigma}^{\nu}-\sigma^{\nu}\bar{\sigma}^{\mu}),\enspace
    \bar{\sigma}^{\mu\nu}=\frac{1}{4}(\bar{\sigma}^{\mu}\sigma^{\nu}-\bar{\sigma}^{\nu}\sigma^{\mu}).
\end{equation}
Accordingly, the corresponding generators of conformal transformations in four dimensions are
\begin{equation}\label{G_Q}
    \begin{aligned}
    P_{\mu}=\left(\begin{array}{cc}
        0 & -h{\sigma}_{\mu}\\
        0 & 0
        \end{array}\right),\enspace & 
    K_{\mu}=\left(\begin{array}{cc}
        0 & 0\\
        h\bar{\sigma}_{\mu} & 0
        \end{array}\right),\\
    M_{\mu\nu}=-h\left(\begin{array}{cc}
        \sigma_{\mu\nu} & 0\\
        0 & \bar{\sigma}_{\mu\nu}
        \end{array}\right),\enspace & 
    D=-\frac{h}{2}\left(\begin{array}{cc}
         1 & 0\\
        0 & -1
        \end{array}\right).
    \end{aligned}
\end{equation}
And their commutation relations can be easily examined, 
\begin{equation}
    \begin{aligned}
        [D,P_{\mu}]= & -hP_{\mu},\\
        [D,K_{\mu}]= & hK_{\mu},\\
        [K_{\mu},P_{\nu}]= & -2h(\eta_{\mu\nu}D-M_{\mu\nu}),\\
        [K_{\rho},M_{\mu\nu}]= & -h(\eta_{\rho\mu}K_{\nu}-\eta_{\rho\nu}K_{\mu}),\\
        [P_{\rho},M_{\mu\nu}]= & -h(\eta_{\rho\mu}P_{\nu}-\eta_{\rho\nu}P_{\mu}),\\
        [M_{\mu\nu},M_{\rho\sigma}]= & -h(\eta_{\mu\sigma}M_{\nu\rho}+\eta_{\nu\rho}M_{\mu\sigma}\\
         & -\eta_{\mu\rho}M_{\nu\sigma}-\eta_{\nu\sigma}M_{\mu\rho}).
    \end{aligned}
\end{equation}
And their Hermitian conditions \cite{32} are
\begin{equation}\label{H_G}
    \begin{aligned}
        P_{0}^{\dagger}=K_{0},\enspace&P_{i}^{\dagger}=-K_{i},\\
        K_{0}^{\dagger}=P_{0},\enspace&K_{i}^{\dagger}=-P_{i},\\
        M_{0i}^{\dagger}=-M_{0i},\enspace&M_{ij}^{\dagger}=M_{ij},\\
        D^{\dagger}=&-D,
    \end{aligned}
\end{equation}
where the Hermitian conjugate \cite{24} of biquaternion matrix $\Lambda_{ab}$ is defined to be 
\begin{equation}
    (\Lambda_{ab})^\dagger=\Lambda^\dagger_{ba}=\bar{\Lambda}^\ast_{ba},
\end{equation}
in which $a$ and $b$ represent the matrix's row and column indices, respectively. Our Hermitian condition \eqref{H_G} is equivalent to that in radial quantization \cite{32}. 

To apply conformal transformations Eqs. (\ref{Tral}, \ref{SCT}, \ref{R_D}), to fields in physics, we use the Pauli matrices as basis of biquaternion
\begin{equation}\label{cond}
    h=-i,\enspace \boldsymbol{i}=-i\sigma_1,\enspace \boldsymbol{j}=-i\sigma_2,\enspace \boldsymbol{k}=-i\sigma_3,
\end{equation} 
where Pauli matrices $\sigma_i$ ($i=1,2,3$) are defined to be
\begin{eqnarray*}
    \sigma_{1}=\left(\begin{array}{cc}
    0 & 1\\
    1 & 0
    \end{array}\right), \enspace & \sigma_{2}=\left(\begin{array}{cc}
    0 & -i\\
    i & 0
    \end{array}\right), \enspace & \sigma_{3}=\left(\begin{array}{cc}
    1 & 0\\
    0 & -1
    \end{array}\right).
\end{eqnarray*}
Substituting this condition \eqref{cond} into Eq. \eqref{G_Q}, we can get
\begin{equation}\label{G_S}
    \begin{aligned}
        P_{\mu}=i\left(\begin{array}{cc}
            0 & {\sigma}_{\mu}\\
            0 & 0
            \end{array}\right),\enspace & 
        K_{\mu}=-i\left(\begin{array}{cc}
            0 & 0\\
            \bar{\sigma}_{\mu} & 0
            \end{array}\right),\\
        M_{\mu\nu}=i\left(\begin{array}{cc}
            \sigma_{\mu\nu} & 0\\
            0 & \bar{\sigma}_{\mu\nu}
            \end{array}\right),\enspace & 
        D=\frac{i}{2}\left(\begin{array}{cc}
             1 & 0\\
            0 & -1
            \end{array}\right).
    \end{aligned}
\end{equation}
Using the Dirac matrices, the generators \eqref{G_S} are rewritted as
\begin{subequations}
    \begin{align}
        P_{\mu}= & \frac{i}{2}\gamma_{\mu}(1+\gamma_{5}),\\
        M_{\mu\nu}= & \frac{i}{2} \gamma_{\mu\nu}=\frac{i}{4}[\gamma_{\mu},\gamma_{\nu}]\\
        D= & -\frac{i}{2}\gamma_{5},\\
        K_{\mu}= & -\frac{i}{2}\gamma_{\mu}(1-\gamma_{5}),
    \end{align}
\end{subequations}
here Dirac matrices are defined to be in (chiral) Weyl representation
\begin{eqnarray*}
    \gamma^{0}=\left(\begin{array}{cc}
    0 & 1\\
    1 & 0
    \end{array}\right), \enspace
    & \gamma^{i}=\left(\begin{array}{cc}
    0 & \sigma_{i}\\
    -\sigma_{i} & 0
    \end{array}\right), \enspace
    & \gamma^{5}=\left(\begin{array}{cc}
    -1 & 0\\
    0 & 1
    \end{array}\right).
\end{eqnarray*}
Clearly by such a process we have derived the spinor representation of the conformal group \cite{17,18,19}.

\subsection{Associating inversions of Spinor Representation with discrete 
 spacetime transformations}

In addition, while we have obtained the above spinor representations of finite conformal transformations, we find out that under the condition $t^2-x^2-y^2-z^2=1$, inversions could be related with discrete spacetime transformations. We shall first establish the connection of and temporal inversion \eqref{I_0} with parity $P$. Substituting this matrix \eqref{I_0} into Eq. \eqref{Def_LFT_Q}, we obtain
\begin{equation}\label{Parity_Q}
    \begin{aligned}
        q^{\prime}= & (aq+b)(cq+d)^{-1}\\
        = & (t+x(hi)+y(hj)+z(hk))^{-1}\\
        = & ((t-x(hi)-y(hj)-z(hk))\\
          & (t+x(hi)+y(hj)+z(hk)))^{-1}\\
          & (t-x(hi)-y(hj)-z(hk))\\
        = & (t^{2}-x^{2}-y^{2}-z^{2})^{-1}\\
          & (t-x(hi)-y(hj)-z(hk))\\
        = & t-x(hi)-y(hj)-z(hk),  
    \end{aligned}
\end{equation}
where we set $t^2-x^2-y^2-z^2=1$ henceforth. The corresponding spacetime transformation is
\begin{equation}
    t^{\prime}=t,\enspace x^{\prime}=-x,\enspace y^{\prime}=-y,\enspace z^{\prime}=-z.
\end{equation}
Here the above temporal inversion corresponds to the spacetime transformation as follows  
\begin{equation}
    x^{\prime}_{I_0}= (t,-x,-y,-z).
\end{equation}
Obviously, this spacetime transformation can be viewed as Partity. According to above definition of Dirac matrices, we find out that $\gamma^0$ in the Weyl representation plays the role of a bridge linking both
temporal inversion and parity,
\begin{equation}
     \Lambda_{I_{0}}= \left(\begin{array}{cc}
        0 & 1\\
        1 & 0
        \end{array}\right)=\gamma^{0}. 
\end{equation}
Consequently, under parity transformation the Dirac spinors change as follows
\begin{equation}\label{parity}
    \begin{aligned}
        P\psi(x)P= & \Lambda_{I_0}\psi(\Lambda_{I_0}^{-1}x)\\
= & \gamma^{0}\psi(t,-\boldsymbol{x}).
    \end{aligned}
\end{equation}

Now let us discuss the relationship of spatial inversions \eqref{I_i} and time reversal $T$. Similar to Eq. \eqref{Parity_Q}, substituting matrices \eqref{I_i} into Eq. \eqref{Def_LFT_Q}, one finds out that the spatial inversions correspond to the following spacetime transformations,
\begin{equation}\label{I_i_con}
    \begin{aligned}
        x^{\prime}_{I_1}=& (-t,x,-y,-z),\\
        x^{\prime}_{I_2}=& (-t,-x,y,-z),\\
        x^{\prime}_{I_3}=& (-t,-x,-y,z).
    \end{aligned}
\end{equation}
Then, substituting the condition \eqref{cond} into Eqs. \eqref{I_i}, we obtain
\begin{equation}
    \Lambda_{I_{i}}=  \left(\begin{array}{cc}
        0 & -i\sigma_{i}\\
        i\sigma_{i} & 0
        \end{array}\right)=-i\gamma^{i}.  
\end{equation}
Eq. \eqref{I_i_con} suggests us to regard following product $(-i\gamma^1)
(-i\gamma^2)(-i\gamma^3)$ as the time reversal,  
\begin{equation}
    x^{\prime}_{T}=x^{\prime}_{I_1 I_2 I_3}= (-t,x,y,z).
\end{equation} 
Based on such mapping we define
\begin{equation}
    \begin{aligned}
        \Lambda_{T}= & (-i\gamma^1)(-i\gamma^2)(-i\gamma^3)\\
        = & i\gamma^1\gamma^2\gamma^3=
          \left(\begin{array}{cc}
          0 & 1\\
          -1 & 0
          \end{array}\right).
    \end{aligned}
\end{equation}
Physically, the time reversal shall be responsible for switch particle to antiparticle due to the Feynman picture of an antiparticle motion just as a particle moving backwards in time \cite{33,34}. Under time reversal transformation, the Dirac spinors transform as follows
\begin{equation}\label{T}
    \begin{aligned}
        T\psi(x)T= & \Lambda_{T}\psi^c(\Lambda_{T}^{-1}x)\\
        = & i\gamma^{1}\gamma^{2}\gamma^{3}\psi^c(-t,\boldsymbol{x})\\
        = & \gamma^{1}\gamma^{3}\psi^{\ast}(-t,\boldsymbol{x}),
    \end{aligned}    
\end{equation}
where $\psi^c$ represents the transformed Dirac spinor by conventional charge conjugation operator $C$, as follows 
\begin{equation}\label{c}
    \psi^c=C\psi C=-i\gamma^2 \psi^{\ast},    
\end{equation}
according to which we have indirectly obtained the Dirac-Matrix form of charge conjugation \cite{26}. The above Eqs. (\ref{parity}, \ref{T}, \ref{c}) illustrate how inversions relate to discrete spacetime transformations, among which Eqs. \eqref{parity} and \eqref{c} are the essential ones we shall apply in the next section. We stress that condition $t^2-x^2-y^2-z^2=1$ does not affect the nonlinearity in Eq. \eqref{Jac} as well as the nonlinear analysis in the next section.

\section{Perform conformal transformations on Dirac Spinors}\label{Sec_4}

To track the nonlinear effects of translations and SCTs, we shall perform them on Dirac spinors.  
However, these transformations exceed the traditional sense of those pertaining Poincar\'e group, in contrast, they only occur maybe in accelerating environments according to their physical understanding \cite{09}. We thus suppose they could be applied to the primordial soup of particles at a moment of very early universe \cite{35}. Specifically, we apply the  transformations to Yukawa term at the moment electroweak symmetry began to break, then to vector-spinor interaction. Subsequently we check the variations of $CP$ symmetry in the transformed Yukawa term and vector-spinor interaction.

Now we give a short explanation why the SCT has the physical meaning of
acceleration. Following its original definition \eqref{Def_CT}, we will reduce SCTs through a series of conditions \cite{09}. If we set $b^\mu=(0,\frac{1}{2}\boldsymbol{g})$ and with $x^\mu=(t,\boldsymbol{x})$, SCTs can be reduced to
\begin{equation}
    \begin{aligned}
        t^{\prime}= & \frac{t}{1+\boldsymbol{g}\cdot\boldsymbol{x}-\frac{1}{4}\boldsymbol{g^{2}}(t^{2}-\boldsymbol{x^{2}})},\\
        \boldsymbol{x^{\prime}}= & \frac{\boldsymbol{x}-\frac{1}{2}\boldsymbol{g}(t^{2}-\boldsymbol{x^{2}})}{1+\boldsymbol{g}\cdot\boldsymbol{x}-\frac{1}{4}\boldsymbol{g^{2}}(t^{2}-\boldsymbol{x^{2}})}.
    \end{aligned}
\end{equation}
At spatial coordinate origin $\boldsymbol{x}=0$, we can obtain
\begin{equation}
    t^{\prime}= \frac{t}{1-\frac{1}{4}\boldsymbol{g^{2}}t^{2}},\enspace
    \boldsymbol{x^{\prime}}=  \frac{-\frac{1}{2}\boldsymbol{g}t^{2}}{1-\frac{1}{4}\boldsymbol{g^{2}}t^{2}}.
\end{equation}
When we consider infinitesimal transformation $\boldsymbol{g}\rightarrow 0$, the SCT can be reduced to 
\begin{equation}\label{acc}
    t^{\prime}= t,\enspace \boldsymbol{x^{\prime}}=  -\frac{1}{2}\boldsymbol{g}t^{2}.
\end{equation}
From Eq. \eqref{acc}, SCTs reflect accelerated motion at a certain moment. Furthermore we notice that generators $P_\mu$ and $K_\mu$ are symmetric among the commutation relations, thus they are on an equal footing in the
conformal group. We define them uniformly as
physical translation. In quantum physics, there is no conception of
acceleration. Meanwhile in the development of quantum theory, we are used to
the transformation of translation, which was supposed to happen instantly, without acceleration at the beginning moment and ending moment. Combining their physics meaning together, here we draw the ideal definitions of translations and SCTs back to reality, i.e. they should be true physical translations in quantum sense. 

Now we consider the transformations of interaction involving scalar field $\phi$ (maybe
viewed as certain deformed Higgs field), at the moment electroweak symmetry just breaking. The related Yukawa term \cite{33} is 
\begin{equation}
    \mathcal{L}_{\text{Yuk}}=g\bar{\psi}\psi\phi, 
\end{equation}
where $g$ is coupling constant. Henceforth the transformations we consider involve only those spinor parts without coordinate transformations, which has been thoroughly studied \cite{06}. Under conformal transformations, the Dirac spinors change as follows 
\begin{equation}
    \begin{aligned}
        \psi^\prime=&D(\Lambda)\psi,\\
        \psi^{\dagger\prime}=&(D(\Lambda)\psi)^{\dagger}=\psi^{\dagger}D^{\dagger}(\Lambda),
    \end{aligned}
\end{equation}
where $D(\Lambda)=\Lambda_P$, $\Lambda_K$, $\Lambda_D$, $\Lambda_M$ are conformal transformations as Eqs. (\ref{Tral}, \ref{SCT}, \ref{R_D}),  acting on spinor indices. Under conformal transformations, the Yukawa term becomes 
\begin{equation}
    \mathcal{L}_{\text{Yuk}}^\prime=g\psi^{\dagger}D^{\dagger}(\Lambda)\gamma^{0} D(\Lambda)\psi \phi .
\end{equation}
Using the Dirac matrices, the translations \eqref{Tral} and SCTs \eqref{SCT} are rewritted as
\begin{align}
    \label{Tral_S} \Lambda_{P}=&1+\frac{1}{2}a^{\mu}\gamma_{\mu}(1+\gamma_{5}),\\
    \label{SCT_S} \Lambda_{K}=&1-\frac{1}{2}b^{\mu}\gamma_{\mu}(1-\gamma_{5}).
\end{align}
Using Eq. \eqref{SCT_S}, the Yukawa term under SCTs becomes
\begin{equation}\label{Y_SCT}
    \begin{aligned}
    (\mathcal{L}_{\text{Yuk}}^{\prime})_{\text{SCT}}=&g\psi^{\dagger}\Lambda_{K}^{\dagger}\gamma^{0}\Lambda_{K}\psi\phi=g\psi^{\dagger}\gamma^{0}\Lambda_{K}^{2}\psi\phi\\=&g\psi^{\dagger}\gamma^{0}(1-b^{\mu}\gamma_{\mu}(1-\gamma^{5}))\psi\phi.
    \end{aligned}
\end{equation} 
Similarly the Yukawa term under translations becomes
\begin{equation}\label{Y_Tral}
     \begin{aligned}
         (\mathcal{L}_{\text{Yuk}}^{\prime})_{\text{translation}}= & g\psi^{\dagger}\Lambda_{P}^{\dagger}\gamma^{0}\Lambda_{P}\psi\phi=g\psi^{\dagger}\gamma^{0}\Lambda_{P}^{2}\psi\phi\\
= & g\psi^{\dagger}\gamma^{0}(1+a^{\mu}\gamma_{\mu}(1+\gamma^{5}))\psi\phi.
     \end{aligned}
\end{equation}
Comparing Eqs. \eqref{Y_SCT} and \eqref{Y_Tral} with Eq. \eqref{Jac}, we find out that finite term $b^2 x^2$ does not appear in Eqs. \eqref{Y_SCT} and \eqref{Y_Tral}. So the translations \eqref{Y_SCT} and \eqref{Y_Tral} turn out to be linear for the Yukawa term.
Next, let's consider the effect of SCTs on $CP$ symmetry in the Yukawa term. Using Eqs. (\ref{parity}, \ref{c}, \ref{Y_SCT}), we can get  
\begin{equation}\label{CPY}
    \begin{aligned} 
        & CP(\mathcal{L}_{\text{Yuk}}^{\prime})_{\text{SCT}}CP\\
= & CPg\psi^{\dagger}\gamma^{0}(1-b^{\mu}\gamma_{\mu}(1-\gamma^{5}))\psi\phi CP\\
= & g\psi^{\dagger}\gamma^{0}(+1-(-1)^{\rho}(-b^{\mu}\gamma_{\mu}(1-\gamma^{5})))\psi\phi,
    \end{aligned}
\end{equation} 
where we use the common shorthand $(-1)^\mu \equiv 1$ for $\mu=0$ and  $(-1)^\mu \equiv -1$ for $\mu=1,2,3$.
Surprisingly, the transformed Yukawa term \eqref{Y_SCT} is not $CP$ invariant (translations similar to SCTs), which means physical translations may actually contribute to $CP$ asymmetry. 

After electroweak symmetry breaking, it naturally reminds us to check further the transformations of the vector-spinor interaction undergo conformal transformations. The vector-spinor interaction can be expressed as
\begin{equation}
    \mathcal{L}_{\text{I}}=g\bar{\psi}\gamma^\mu\psi A_\mu.
\end{equation} 
Under conformal transformations, the vector-spinor interaction becomes 
\begin{equation}
    \mathcal{L}_{\text{I}}^{\prime}=g\psi^{\dagger}D^{\dagger}(\Lambda)\gamma^{0}\gamma^{\mu}D(\Lambda)\psi A_{\mu}.
\end{equation}
Under SCTs \eqref{SCT_S}, the vector-spinor interaction becomes
\begin{equation}\label{Int_SCT}
    \begin{aligned}
        (\mathcal{L}_{\text{I}}^{\prime})_{\text{SCT}}= & g\psi^{\dagger}\Lambda_{K}^{\dagger}\gamma^{0}\gamma^{\mu}\Lambda_{K}\psi A_{\mu}\\
= & g\psi^{\dagger}\gamma^{0}(\gamma^{\mu}-b_{\rho}(g^{\rho\mu}+4DM^{\rho\mu})\Lambda_{K})\psi A_{\mu}\\
= & g\psi^{\dagger}\gamma^{0}(\gamma^{\mu}-b_{\rho}(g^{\mu\rho}+\gamma^{5}\gamma^{\rho\mu})\\
 & +\frac{1}{2}b^{\mu}b^{\sigma}\gamma_{\sigma}(1-\gamma^{5})\\
 &-\frac{1}{2}b_{\rho}b^{\sigma}\gamma^{\rho\mu}\gamma_{\sigma}(1-\gamma^{5}))\psi A_{\mu}.  
    \end{aligned}
\end{equation}
When compared with Eq. \eqref{Jac} it reveals that the properties of SCTs manifest a combination of infinitesimal dilation and rotation. The vector-spinor interaction under translations \eqref{Tral_S} becomes
\begin{equation}\label{Int_Tral}
    \begin{aligned}
       & (\mathcal{L}_{\text{I}}^{\prime})_{\text{translation}}\\= & g\psi^{\dagger}\Lambda_{P}^{\dagger}\gamma^{0}\gamma^{\mu}\Lambda_{P}\psi A_{\mu}\\
= & g\psi^{\dagger}\gamma^{0}(\gamma^{\mu}+a_{\rho}(g^{\rho\mu}-4DM^{\rho\mu})\Lambda_{P})\psi A_{\mu}\\=  & g\psi^{\dagger}\gamma^{0}(\gamma^{\mu}+a_{\rho}(g^{\mu\rho}-\gamma^{5}\gamma^{\rho\mu})\\
 & +\frac{1}{2}a^{\mu}a^{\sigma}\gamma_{\sigma}(1+\gamma_{5})\\
 & +\frac{1}{2}a_{\rho}a^{\sigma}\gamma^{\rho\mu}\gamma_{\sigma}(1+\gamma_{5}))\psi A_{\mu}.
    \end{aligned}
\end{equation}
In Eqs. \eqref{Int_SCT} and \eqref{Int_Tral}, finite term can be reserved. In other words, the nonlinearity may be manifested. 
Next, we consider the effect of SCTs on $CP$ symmetry in the vector-spinor interaction. Using Eqs. (\ref{parity}, \ref{c}, \ref{Int_SCT}), we can get
\begin{equation}\label{CPI}
    \begin{aligned}
          & CP(\mathcal{L}_{\text{I}}^{\prime})_{\text{SCT}}CP\\
= & CPg\psi^{\dagger}\gamma^{0}(\gamma^{\mu}-b^{\rho}(g^{\mu\rho}+\gamma^{\rho\mu}\gamma^{5})\\
 & +\frac{1}{2}b^{\mu}b_{\sigma}\gamma^{\sigma}(1-\gamma^{5})\\
 & -\frac{1}{2}b_{\rho}b_{\sigma}\gamma^{\rho\mu}\gamma^{\sigma}(1-\gamma^{5}))\psi A_{\mu}CP\\
= & g\psi^{\dagger}\gamma^{0}(+\gamma^{\mu}-(-1)^{\rho}(-b^{\rho}(g^{\mu\rho}+\gamma^{5}\gamma^{\rho\mu}))\\
 & +(-1)^{\sigma}(-1)^{\mu}\frac{1}{2}b^{\mu}b^{\sigma}\gamma_{\sigma}(1-\gamma^{5})\\
 & \pm(-1)^{\rho}(-1)^{\sigma}(-\frac{1}{2}b_{\rho}b^{\sigma}\gamma^{\rho\mu}\gamma_{\sigma}(1-\gamma^{5})))\psi A_{\mu},
    \end{aligned}
\end{equation}
where we use the shorthand  $\pm \equiv +1$ for $\sigma=\rho$, $\sigma \neq \mu$ or $\sigma\neq\rho$, $\sigma = \mu$ and $\pm \equiv -1$ for $\sigma=\rho=\mu$ or $\sigma\neq \rho$, $\sigma \neq \mu$.
In summary, the transformed vector-spinor interaction \eqref{Int_SCT} is not $CP$ invariant.

\section{Discussions}\label{Sec_5}
So far we have clarified the relationship between spacetime transformations and spinor representations of four-dimensional conformal group, by replacing complex numbers with biquaternions in linear fractional transformations. The work endows deeper insight in how Conformal Transformations \cite{36} closely relate with Spinor Theory \cite{20} in an intuitive manner. Both Conformal Representation and Spinor Theory (Clifford Algebra) have their own geometric origins, but the two geometries are totally different. At first sight one cannot see through that spinor theory somehow relates to isogonal transformation. In a previous work \cite{17}, we have plausibly constructed the relationship between four-basis Clifford Algebra and Conformal transformations, but there we confirmed the components of spinor matrices to be generators of conformal group merely by their commutation relations. That seems a rigid mapping, omitting both of their geometric backgrounds. Here to unify their geometric backgrounds, we thus tried to derive the spinor matrices from linear fractional transformation. From the construction of spinor \cite{20} we saw its connection with Clifford algebra and quaternion, that suggested us to employ quaternion as tentative tool, latter we find only biquaternion works for the derivation. 

The conventional method of extending conformal transformation from two dimension to higher dimension via metric tensor is straightforward, however on one hand that conceals the nonlinearity of conformal transformation. On the other hand that conceals the critical component, i.e. inversion, of conformal transformations. Through the manner of this paper we reveal the two respects by showing their spacetime effects in spinor representation. As an example we investigate how the spacetime inversion corresponds to parity and time reversal in spinor representation, under the condition $t^2-x^2-y^2-z^2=1$ and physical requirement. Finally we find out that finite conformal transformations are associated with a special set of spinor matrices, which turns out to be Weyl representation. Furthermore the Hermitian conditions for generators of the conformal group are derived from Hermitian properties of biquaternions.  The above results are qualitatively coincident with those relevant respects appeared in literatures \cite{18,26,32,33,34,37,38}. 

The nonlinearity of the conformal group could be manifested via the vector-spinor interaction  by considering the physical translation (we have defined translations and SCTs as physical translation in Section I) of Dirac spinors. And such nonlinearity leads to important subsequences in the very early moment of our universe, just after the Big Bang. Conditionally, one of the subsequences is $CP$ violation at the moment when electroweak symmetry breaking. From Eqs. \eqref{CPY} and \eqref{CPI}, if the parameter $b^\mu$ transforms as a vector field like  $A^\mu$, then the $CP$ violation of Yukawa term would not occur. As for the vector-spinor interaction, however, the $CP$ violation is free of this condition and meanwhile accompanied with nonlinearity. So far we have attributed the symmetry breaking arising from SCTs to certain acceleration, which is caused by the above-mentioned physical translations. It is obvious here that physical translation might correspond to the effects of classical acceleration \eqref{acc}, whereas in quantum physics there is no conception of acceleration. This cognition of physical translation may open new avenues for exploring large-scale universe, for instance in studying dark matter, dark energy, cosmology neutrino physics, gamma-ray burst, etc.

\section{Acknowledgment}

 We thank Jiyuan Ke for his helpful suggestions. 

\bibliography{Ref}

\end{document}